\documentclass
[aps,twocolumn,showpacs,superscriptaddress,tightenlines]{revtex4}%
\usepackage{epsfig,rotating}
\usepackage{amsmath}
\usepackage{amsfonts}
\usepackage{amssymb}
\usepackage{graphicx}%
\begin{document}
\title{Relativistic continuum-continuum coupling in the dissociation of halo nuclei}
\author{C.\,A.~Bertulani}
\email{bertulani@physics.arizona.edu}
\affiliation{Department of Physics, University of Arizona, Tucson, Arizona 85721}
\date{\today}

\begin{abstract}
A relativistic coupled-channels theory for the calculation of
dissociation cross sections of halo nuclei is developed. A
comparison with non-relativistic models is done for the dissociation
of $^{8}$B projectiles. It is shown that \ neglecting relativistic
effects leads to seizable inaccuracies in the extraction of the
astrophysical S-factor for the proton+beryllium radiative capture
reaction.

\end{abstract}
\maketitle


\label{intr} Reactions with radioactive nuclear beams have quickly
become a major research area in nuclear physics. Among the newly
developed techniques, the Coulomb dissociation method is an
important tool to obtain electromagnetic matrix elements between
continuum and bound states of rare nuclear isotopes \cite{BBH86}.
These matrix elements play an essential role in nuclear
astrophysics. At low continuum energies, they are the same as the
ones involved in radiative capture processes of astrophysical
interest. In particular, the Coulomb dissociation of $^{8}$B
projectiles allows to extract information on the radiative capture
reaction p$+^{7}$Be $\rightarrow$\ $^{8}$B$+\gamma,$ of relevance
for the standard solar model and the production of high-energy
neutrinos in the sun \cite{RR88}.

The dissociation of weakly-bound nuclei, or halo nuclei, is
dominated by the Coulomb interaction, although the nuclear
interaction with the target cannot be neglected in most cases. The
final state interaction of the fragments with the target leads to
important continuum-continuum and continuum-bound-state couplings
which appreciably modify the reaction dynamics. Higher-order
couplings are more relevant in the dissociation of halo nuclei due
to their low binding. A known example is the \textquotedblleft
post-acceleration\textquotedblright\ (or \textquotedblleft
reacceleration\textquotedblright) effect observed in the
dissociation of $^{11}$Li projectiles \cite{BC92,BBK92,Iek93,BB93}.

Two methods have been devised to study higher-order effects in
projectile dissociation. The method introduced in ref. \cite{BB93}
uses the direct solution of the Schr\"{o}dinger equation (DSSE) by
space-time discretization. One starts with a ground-state
wavefunction, propagates it through each time-step, and after
sufficiently long time the iterated wavefunction is projected into a
specific channel of interest. Another method discretizes the
continuum wavefunctions $\left\vert c\right\rangle $\ and uses them
as input to calculate the matrix elements $\left\langle c^{\prime
}\left\vert V_{int}\right\vert c\right\rangle $ and $\left\langle
c\left\vert V_{int}\right\vert b\right\rangle $, where $\left\vert
b\right\rangle $ denotes a bound state \cite{BC92}. The matrix
elements are then used in a coupled-channels calculation for
transition amplitudes to dissociation channels. This is known as
continuum discretized coupled-channels (CDCC) method and was
introduced by Rawitscher \cite{Ra74} to study nuclear breakup
reactions of the type $a+A\ \rightarrow b+c+A$. It has been used
extensively in the study of breakup reactions with stable
\cite{SYK86} and unstable nuclear projectiles \cite{BC92,NT98,NT99}

Special relativity, an obviously important concept in physics, is
quite often neglected in the afore mentioned dynamical calculations.
Most rare isotope facilities use projectile dissociation at 100
MeV/nucleon. At these energies, relativistic effects are expected to
be of the order of 10\%. Relativistic effects are accounted for in
the collision kinematics, in first-order perturbation calculations,
but not in dynamical calculations used sofar in the analysis of
experiments. They also enter the dynamics in a non-linear, often
unpredictable, way. The reason why these effects have not been
considered before is due to the inherent theoretical difficulty in
defining a nuclear potential between many-body relativistic systems.
Due to retardation, an attempt to use a microscopic description
starting from binary collisions of the constituents is not possible.
A successful approach, known as \textquotedblleft Dirac
phenomenology\textquotedblright, has been achieved for
nucleon-nucleus scattering \cite{AC79}. But for nucleus-nucleus
collisions a reasonable account of these features has not yet been
accomplished. In the present letter, the case of dissociation of
$^{8}$B projectiles is studied. The major contribution comes from
the Coulomb interaction with well-known relativistic transformation
properties. A coupled-channels method based on the eikonal
approximation with relativistic ingredients is developed and
compared to semiclassical methods \cite{BB93}. Here I omit the
consideration of other corrections in the eikonal treatment of the
scattering involving halo nuclei which have also been shown to play
a relevant role (see, e.g., refs. \cite{Al97,EB99}).

Let us consider the Klein-Gordon (KG) equation with a potential $V_{0}$ which
transforms as the time-like component of a four-vector \cite{AC79}. For a
system with total energy $E$ (including the rest mass $M$), the KG equation
can be cast into the form of a Schr\"{o}dinger equation (with $\hbar=c=1$),
$\left(  \nabla^{2}+k^{2}-U\right)  \Psi=0,$ where $k^{2}=\left(  E^{2}%
-M^{2}\right)  $ and $U=V_{0}(2E-V_{0})$. When $V_{0}\ll M$, and
$E\simeq M$, one gets $U=2MV_{0}$, as in the non-relativistic case.
The condition $V_{0}\ll M$ is met in peripheral collisions between
nuclei at all collision energies. Thus, one can always write
$U=2EV_{0}$. A further simplification is to assume that the center
of mass motion of the incoming projectile and outgoing fragments is
only weakly modulated by the potential $V_{0}$. To get the dynamical
equations, one discretizes the wavefunction in terms of the
longitudinal center-of-mass momentum $k_{z}$, using the ansatz%
\begin{equation}
\Psi=\sum_{\alpha}\mathcal{S}_{\alpha}\left(  z,\mathbf{b}\right)
\ \exp\left(  ik_{\alpha}z\right)  \ \phi_{k_{\alpha}}\left(
\mathbf{\mbox{\boldmath$\xi$}}\right)  \ . \label{eq1}%
\end{equation}
In this equation, $\left(  z,\mathbf{b}\right)  $ is the
projectile's center-of-mass coordinate, with \textbf{b} equal to the
transverse coordinate. $\ \phi\left(
\mathbf{\mbox{\boldmath$\xi$}}\right) $ is the projectile intrinsic
wavefunction and $\left( k,\mathbf{K}\right) $ is the projectile's
center-of mass momentum with longitudinal momentum $k$\ and
transverse momentum $\mathbf{K}$\textbf{.} There are hidden,
uncomfortable, assumptions in eq. \ref{eq1}. The separation between
the center of mass and intrinsic coordinates is not permissible
under strict relativistic treatments. For high energy collisions we
can at best justify eq. \ref{eq1} for the scattering of light
projectiles on heavy targets. Eq. \ref{eq1} is only reasonable if
the projectile and target closely maintain their integrity during
the collision, as in the case of very peripheral collisions.

Neglecting the internal structure means $\phi_{k_{\alpha}}\left(
\mathbf{\mbox{\boldmath$\xi$}}\right)  =1$ and the sum in eq.
\ref{eq1} reduces to a single term with $\alpha=0$, the projectile
remaining in its ground-state. It is straightforward to show that
inserting eq. \ref{eq1} in the KG equation $\left(
\nabla^{2}+k^{2}-2EV_{0}\right)  \Psi=0$, and neglecting
$\nabla^{2}\mathcal{S}_{0}\left(  z,\mathbf{b}\right)  $ relative to
$ik\partial_{Z}\mathcal{S}_{0}\left(  z,\mathbf{b}\right)  $, one
gets
$ik\partial_{Z}\mathcal{S}_{0}\left(  z,\mathbf{b}\right)  =EV_{0}%
\mathcal{S}_{0}\left(  z,\mathbf{b}\right)  $, which leads to the
center of mass scattering solution $\mathcal{S}_{0}\left(
z,\mathbf{b}\right) =\exp\left[
-iv^{-1}\int_{-\infty}^{z}dz^{\prime}\ V_{0}\left(  z^{\prime
},\mathbf{b}\right)  \right]  ,$ with $v=k/E$. Using this result in
the Lippmann-Schwinger equation, one gets the familiar result for
the eikonal elastic scattering amplitude, i.e. $f_{0}=-i\left( k/
2\pi\right) \int d\mathbf{b}\ \exp\left(  i\mathbf{Q\cdot b}\right)
\ \left\{  \exp\left[ i\chi(\mathbf{b})\right]  -1\right\}  , $
where the eikonal phase is given by
$\exp[i\chi(\mathbf{b})]=\mathcal{S}_{0}\left(
\infty,\mathbf{b}\right) $, and
$\mathbf{Q}=\mathbf{K}^{\prime}-\mathbf{K}$ is the transverse
momentum transfer. \ Therefore, the elastic scattering amplitude in
the eikonal approximation has the same form as that derived from the
Schr\"odinger equation in the non-relativistic case.

For inelastic collisions we insert eq. \ref{eq1} in the KG equation
and use the orthogonality of the intrinsic wavefunctions
$\phi_{k_{\alpha}}\left( \mathbf{\mbox{\boldmath$\xi$}}\right)  $.
This leads to a set of coupled-channels equations for
$\mathcal{S}_{\alpha}$:%
\begin{equation}
\left(  \nabla^{2}+k^{2}\right)
\mathcal{S}_{\alpha}\mathrm{e}^{ik_{\alpha}
  z}=\sum_{\alpha }\left\langle \alpha\left\vert U\right\vert
\alpha^{\prime}\right\rangle \ \mathcal{S}_{\alpha^{\prime}}\
\mathrm{e}^{i
k_{\alpha^{\prime}}  z}, \label{eq2}%
\end{equation}
\bigskip with the notation $\left\vert \alpha\right\rangle =\left\vert
\phi_{k_{\alpha}}\right\rangle $. Neglecting terms of the form $\nabla
^{2}\mathcal{S}_{\alpha}\left(  z,\mathbf{b}\right)  $ relative to
$ik\partial_{Z}\mathcal{S}_{\alpha}\left(  z,\mathbf{b}\right)  $, eq.
\ref{eq2} reduces to%
\begin{equation}
iv\frac{\partial\mathcal{S}_{\alpha}\left(  z,\mathbf{b}\right)  }{\partial
z}=\sum_{\alpha^{\prime}}\left\langle \alpha\left\vert V_{0}\right\vert
\alpha^{\prime}\right\rangle \ \mathcal{S}_{\alpha^{\prime}}\left(
z,\mathbf{b}\right)  \ \mathrm{e}^{i\left(  k_{\alpha^{\prime}}-k_{\alpha%
}\right)  z}. \label{eq3}%
\end{equation}
The scattering amplitude for the transition $0\rightarrow\alpha$ is given by%
\begin{equation}
f_{\alpha}\left(  \mathbf{Q}\right)  =-\frac{ik}{2\pi}\int d\mathbf{b}%
\ \ \exp\left(  i\mathbf{Q\cdot b}\right)  \ \left[  S_{\alpha}\left(
\mathbf{b}\right)  -\delta_{\alpha,0}\right]  , \label{eq4}%
\end{equation}
with $S_{\alpha}\left(  \mathbf{b}\right)
=\mathcal{S}_{\alpha}\left( z=\infty,\mathbf{b}\right)  $. The set
of equations \ref{eq3} and \ref{eq4} are the relativistic-CDCC
equations (RCDCC).

I have used the RCDCC equations to study the dissociation of $^{8}$B
projectiles at high energies. The energies transferred to the
projectile are small, so that the wavefunctions can be treated
non-relativistically in the projectile frame of reference. In this
frame the wavefunctions will be described in spherical coordinates,
i.e. $\left\vert \alpha\right\rangle =\left\vert jlJM\right\rangle
$, where $j$, $l$, $J$ and $M$ denote the angular momentum numbers
characterizing the projectile state. Eq. \ref{eq3} is Lorentz
invariant if the potential $V_0$ transforms as the time-like
component of a four-vector. The matrix element $\left\langle
\alpha\left\vert V_0\right\vert \alpha^{\prime}\right\rangle $ is
also Lorentz invariant, and we can therefore calculate them in the
projectile frame.

The longitudinal wavenumber $k_{\alpha}\simeq(E^{2}-M^{2})^{1/2}$\
also defines how much energy is gone into projectile excitation,
since for small energy and momentum transfers
$k_{\alpha}^{\prime}-k_{\alpha}=\left(
E_{\alpha}^{\prime}-E_{\alpha}\right) /v$. In this limit, eqs.
\ref{eq3} and \ref{eq4} reduce to semiclassical coupled-channels
equations, if one uses $z=vt$ for a projectile moving along a
straight-line classical trajectory, and changing to the notation
$\mathcal{S}_{\alpha}\left(  z,b\right) =a_{\alpha }(t,b)$, where
$a_{\alpha}(t,b)$ is the time-dependent excitation amplitude for a
collision wit impact parameter $b$ (see eqs. 41 and 76 of ref.
\cite{BCG03}). Here I use the full version of eq. \ref{eq4}.

If the state $\left\vert \alpha\right\rangle $\ is in the continuum (positive
proton+$^{7}$Be energy) the wavefunction is discretized according to
$\left\vert \alpha;E_{\alpha}\right\rangle =\int dE_{\alpha}^{\prime}%
\ \Gamma(E_{\alpha}^{\prime})\ \left\vert \alpha;E_{\alpha}^{\prime
}\right\rangle $, where the functions $\Gamma(E_{\alpha})$ are
assumed to be strongly peaked around the energy $E_{\alpha}$ with
width $\Delta E$. For convenience the histogram set (eq. 3.6 of ref.
\cite{BC92}) is chosen. The inelastic cross section is obtained by
solving the RCDCC equations and using $d\sigma/d\Omega dE_{\alpha
}=\left\vert f_{\alpha}\left( \mathbf{Q}\right)  \right\vert ^{2}\
\Gamma ^{2}(E_{\alpha})$.

The potential $V_{0}$ contains contributions from the nuclear and
the Coulomb interaction. The nuclear potentials are constructed
along traditional lines of non-relativistic theory. The standard
double-folding approximation
$V_{N}^{(aT)}(\mathbf{R})=\int\rho_{a}\left(  \mathbf{r}\right)
v_{0}\left( \mathbf{s}\right)  \rho_{T}\left(
\mathbf{r}^{\prime}\right)  $ is used, where $v_{0}\left(
\mathbf{s}\right)  $\ is the effective nucleon-nucleon interaction,
with $\mathbf{s=r+R-r}^{\prime}$.\ \ The ground-state densities for
the proton, $^{7}$Be ($\rho_{a}$), and Pb target ($\rho_{T}$), are
taken from ref. \cite{Vri87}. The M3Y effective interaction
\cite{BBML77} is used for $v_{0}\left(  \mathbf{s}\right)  $. The
nucleus-nucleus potential is expanded into $l=0,\ 1,\ 2$
multipolarities. These potentials are then transformed as the
time-like component of a four-vector, as described above (see also
ref. \cite{BCG03}). The multipole expansion of the Coulomb
interaction in the projectile frame including retardation and
assuming a straight-line motion has been derived in ref.
\cite{BCG03}. The first term (monopole) of the expansion is the
retarded Lienard-Wiechert potential which does not contribute to the
excitation, but to the center of mass scattering. Due to its long
range, it is hopeless to solve eq. \ref{eq3} with the Coulomb
monopole potential, as $\mathcal{S}_{\alpha}\left(
z,\mathbf{b}\right)  $\ will always diverge. This can be rectified
by using the regularization scheme $S_{\alpha}\left(
\mathbf{b}\right)  \rightarrow\exp\left[  i\left(  2\eta\ln\left(
kb\right) +\chi_{a}\left(  b\right)  \right)  \right]  \
S_{\alpha}\left( \mathbf{b}\right)  $, where $\eta=Z_{P}Z_{T}/\hbar
v$ and the $S_{\alpha }\left(  \mathbf{b}\right)  $\ on the
right-hand side is now calculated without inclusion of the Coulomb
monopole potential in eq. \ref{eq3}. The purely imaginary absorption
phase, $\chi_{a}\left(  b\right)  $, was introduced to account for
absorption at small impact parameters. It has been calculated using
the imaginary part of the \textquotedblleft t$\rho\rho
$\textquotedblright\ interaction \cite{Ra79}, with the $^{8}$B
density calculated with a modified Hartree-Fock model \cite{Sag04}.
The E1 and E2 interactions, taken from ref. \cite{BCG03}  replacing
$vt \rightarrow z$. Explicitly,
\begin{equation}
V_{E1\mu}=\sqrt{\frac{2\pi}{3}}\xi Y_{1\mu}\left(  \mathbf{\hat
{\mbox{\boldmath$\xi$}}}\right)  \frac{\gamma Z_{T}e_1e}{\left(
b^{2}+\gamma ^{2}z^{2}\right)  ^{3/2}}\left\{
\begin{array}
[c]{c}%
\mp b\ \ \mathrm{if}\ \ \ \mu=\pm1\\
\sqrt{2}z\ \ \mathrm{if}\ \ \ \mu=0\ ,
\end{array}
\right.  \label{eq6}%
\end{equation}
for the E1 (electric dipole) field, and%
\begin{align}
V_{E2\mu}  &  =\sqrt{\frac{3\pi}{10}}\xi^{2}Y_{2\mu}\left(
\mathbf{\hat {\mbox{\boldmath$\xi$}}}\right)  \frac{\gamma
Z_{T}e_2e}{\left(  b^{2}+\gamma
^{2}z^{2}\right)  ^{5/2}}\nonumber\\
&  \times\left\{
\begin{array}
[c]{c}%
b^{2}\ \ \ \ \mathrm{if}\ \ \ \mu=\pm2\\
\mp2\gamma^{2}bz\ \ \ \ \mathrm{if}\ \ \ \mu=\pm1\\
\sqrt{2/3}\left(  2\gamma^{2}z^{2}-b^{2}\right)  \ \ \ \ \mathrm{if}%
\ \ \ \mu=0\ .
\end{array}
\right.  \label{eq7}%
\end{align}
for the E2 (electric quadrupole) field, where $e_1={3 \over 8}e$ and
$e_2={53\over 64}e$ are effective charges for $p+^7$Be. For
$\gamma\rightarrow 1$\ these potentials reduce to the
non-relativistic multipole fields (see, e.g., eq. 2 of ref.
\cite{EB02}) in distant collisions.\begin{figure}[t]
\begin{center}
\includegraphics[
height=3.in,
width=2.5in
]{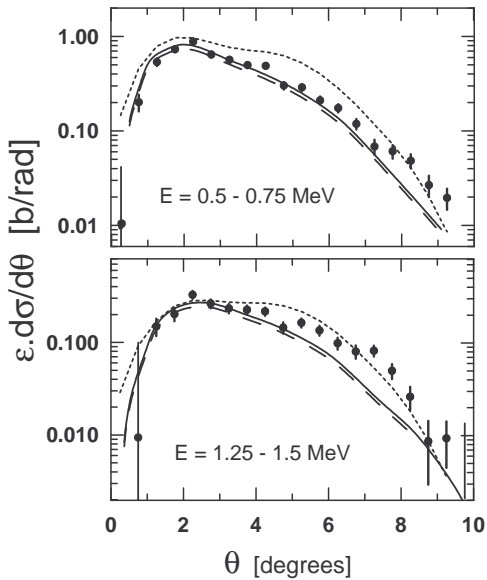}
\end{center}
\caption{ Angular distributions for the dissociation reaction $^{8}$B+Pb
$\rightarrow$ p+$^{7}$Be+Pb at 50 MeV/nucleon. Data are from ref. \cite{Kik97}
and are corrected for the detection efficiency $\varepsilon$. The dotted curve
is the first-order perturbation result of ref. \cite{BG98}. The solid curve is
the RCDCC calculation. The dashed curve is obtained with the replacement of
$\gamma$ by unity in the nuclear and Coulomb potentials.}%
\label{f1}%
\end{figure}

A single-particle model was used for $^{8}$B with a Woods-Saxon
potential adjusted to reproduce the binding energy of 0.139 MeV
\cite{Ro73,EB96,Ber96}. I follow the method of ref. \cite{BC92} and
divide the continuum into bins of widths $\Delta E_{\alpha}=100$
keV, centered at $E_{\alpha}=0.01$, 0.11, 0.21, ..., 1.01 MeV, bins
of widths $\Delta E_{\alpha}=250$ keV, centered at
$E_{\alpha}=1.25,$ 1.5$...,$ 2.0 MeV, and bins of width $\Delta
E_\alpha =0.75$ MeV, centered at $E_\alpha =2.50$, 3.25, ..., 10.00
MeV. Each state $\alpha$ is a combination of energy and angular
momentum quantum numbers $\alpha=\left\{ E_{\alpha },l,j,J,M\right\}
$. Continuum $s$, $p$, $d$ and $f$ waves in $^8$B were included.

The calculations with the RCDCC equations were compared to the data
of refs. \cite{Kik97} and \cite{Dav01}. The results were folded with
the efficiency matrix as well as the energy averaging procedures
explained in ref. \cite{Kik97} and provided by the RIKEN
collaboration \cite{Kik97}. At 83 MeV/nucleon, the angular
distribution was chosen to match the same scattering angles referred
to in ref. \cite{Dav01}.
\begin{figure}[t]
\begin{center}
\includegraphics[
height=2.in,
width=2.9in
]{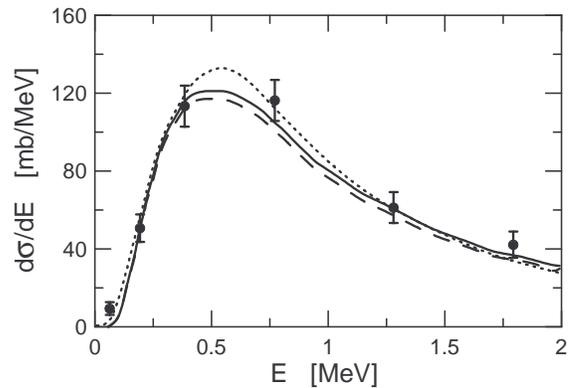}
\end{center}
\caption{ Cross sections for the dissociation reaction $^{8}$B+Pb
$\rightarrow$ p+$^{7}$Be+Pb at 83 MeV/nucleon and for
$\theta_{8}<1.8^{0}$. Data are from ref. \cite{Dav01}. The dotted
curve is the first-order perturbation result. The solid curve is the
RCDCC calculation. The dashed curve is obtained with the replacement
of $\gamma$ by unity in the nuclear and
Coulomb potentials.}%
\label{f2}%
\end{figure}

Figure \ref{f1} shows the angular distributions for the dissociation
reaction $^{8}$B+Pb$\rightarrow$p+$^{7}$Be+Pb at 50 MeV/nucleon.
Data are from ref. \cite{Kik97}. \ The relative energy $E$ between
the proton and $^{7}$Be is averaged over the energy intervals
E=0.5-0.7 MeV (upper panel) and $E=1.25-1.5$ MeV (lower panel). The
dotted curve is the first-order perturbation result reported in ref.
\cite{BG98}. The solid curve is the result of the RCDCC calculation.
The dashed curve is obtained with the replacement of $\gamma$
(Lorentz factor) by unity in the nuclear and Coulomb potentials. The
dashed curve is on average 3-6\% lower than the solid curves in
figure 1. Using non-relativistic potentials yields results always
smaller than the full RCDCC calculation. It is a non-trivial task to
predict what the relativistic corrections do in a coupled-channels
calculation.

Figure 2 shows the relative energy spectrum between the proton and the $^{7}%
$Be after the breakup of $^{8}$B on lead targets at 83 MeV/nucleon.
The data are from ref. \cite{Dav01}. In this case, the calculation
was restricted to $b>30$ fm. The dotted curve is the first-order
perturbation calculation, the solid curve is the RCDCC calculation,
and the dashed curve is obtained with the replacement of $\gamma$ by
unity in the nuclear and Coulomb potentials. The difference between
the solid and the dashed-curve is of the order of 4-9\%. I have also
used the DSSE method, described in ref. \cite{BB93}, to compute the
same spectrum, with the same partial waves, assuming a large lower
cutoff in the parameter of $b=30$ fm. This would justify a
semiclassical limit. The same relativistic nuclear and Coulomb
potentials have been used in both calculations. The difference
between the two results (the RCDCC and the DSSE) is very small (less
than 2\%) for the whole range of the spectrum. To my knowledge such
a comparison has never been made before. This is a proof that the
two methods yield the same result, if the same potentials are used,
and as long as large $b$'s are considered. Such a comparison is only
possible because the same potential was used for the bound-state and
the continuum. The DSSE method \cite{BB93} does not allow for use of
different potentials for p+$^{7}$B. This is not a problem in the
RCDCC method, since the states $\left\vert b\right\rangle $ and
$\left\vert c\right\rangle $ can be calculated within any level of
sophistication, beyond the simple potential model adopted here. In
this respect, the RCDCC is superior than the DSSE method and is more
suitable for an accurate description of dynamical calculations.

The conclusions drawn in this work are crucial in the analysis of
Coulomb breakup experiments at high bombarding energies, as the GSI
experiment at 254 MeV/nucleon \cite{Sch03}. In table 1 I show the
calculations for the correction factor
$\Delta=(\sigma^{RCDCC}-\sigma^{CDCC})/\sigma^{CDCC}$ for the
dissociation of $^8$B on lead targets at 3 bombarding energies. $E$
is the relative energy of the proton and $^7$Be. One sees that the
relativistic corrections tend to increase the cross sections. At 250
MeV/nucleon they can reach a 15\%  value. This has been treated
before in first-order perturbation theory, but not in the dynamical
calculations with continuum-continuum coupling used in the
experimental analysis \cite{Dav01,Sch03}. The consequence of using
these approximations on the extracted values of the astrophysical
S-factors for the $^{7}$Be(p,$\gamma$) reaction in the sun is not
easy to access. It might be necessary to review the results of some
of these data, using a proper treatment of the relativistic
corrections in the theory calculations used in the experimental
analysis. Other improvements of the present formalism needs to be
assessed. The relativistic effects in the nuclear interaction has to
be studied in more depth. The effect of close Coulomb fields
\cite{EB022,EBS04} should also be considered in the case of
dissociation of halo nuclei.

\bigskip

\begin{tabular}
[c]{|l|l|l|l|}\hline
Lab energy & $\Delta$ & $\Delta$ & $\Delta$\\
\lbrack MeV/nucleon] & $E=0.1$ MeV & $E=1$ MeV& $E=2$ MeV\\
\hline 50 & 1.5\% & 4.2\% & 3.4\%\\
\hline 80 & 3\% & 5.5\% & 4.1\%\\
\hline 250 & 5.3\% & 14.6\% & 6.9\%\\
\hline
\end{tabular}

Table 1: Relativistic corrections in the dissociation of  $^8$B
projectiles impinging on lead targets at different bombarding
energies. $E$ is the relative energy of the proton and $^7$Be.

\bigskip\bigskip

I would like to thank T. Aumann, H. Esbensen, K. Suemmerer and I.
Thompson for beneficial discussions. This work was supported by the
U.\thinspace S.\ Department of Energy under grant No.
DE-FG02-04ER41338.


\begin{thebibliography}{99}                                                                                               %


\bibitem {BBH86}G. Baur, C.A. Bertulani and H. Rebel, Nucl. Phys.
\textbf{A458}, 188 (1986).

\bibitem {RR88}C.E. Rolfs, W. Rodney, \textquotedblleft Cauldrons in the
Cosmos\textquotedblright, Chicago Press, 1988.

\bibitem {BC92}C.A.Bertulani and L.F.Canto, Nucl. Phys. \textbf{A540,} 328 (1992).

\bibitem {BBK92}G. Baur, C.A. Bertulani and D.M. Kalassa, Nucl. Phys.
\textbf{A550,} 527 (1992).

\bibitem {Iek93}K. Ieki\textit{\ et al.}, Phys. Rev. Lett. \textbf{70}, 730 (1993).

\bibitem {BB93}G.F. Bertsch and C.A. Bertulani, Nucl. Phys. \textbf{A556,} 136
(1993); C.A. Bertulani and G.F. Bertsch, Phys. Rev. \textbf{C49,}
2839 (1994); H. Esbensen, G.F. Bertsch and C.A. Bertulani, Nucl.
Phys. \textbf{A581,} 107 (1995).

\bibitem {Ra74}G.H. Rawitscher, Phys. Rev. \textbf{C9}, 2210 (1974).

\bibitem {SYK86}Y. Sakuragi, M. Yashiro and M. Kamimura, Prog. Theor. Phys.
Suppl. \textbf{89}, 136 (1986).

\bibitem {NT98}F.M. Nunes and I.J. Thompson, Phys. Rev. \textbf{C57}, R2818 (1998).

\bibitem {NT99}F.M. Nunes and I.J. Thompson, Phys. Rev. \textbf{C59}, 2652 (1999).

\bibitem {AC79}L.G. Arnold and B.C. Clark, Phys. Lett. \textbf{B84}, 46 (1979).

\bibitem {Al97} J.S. Al-Khalili, J.A.Tostevin and J.M. Brooke, Phys.
Rev. {\bf C55}, R1018 (1997).

\bibitem {EB99}H. Esbensen and G.F. Bertsch, Phys. Rev. {\bf C59},
3240 (1999).

\bibitem {BCG03}C.A. Bertulani, C.M. Campbell, and T. Glasmacher, Comput.
Phys. Commun. \textbf{152}, 317 (2003).

\bibitem {Vri87}H. De Vries, C.W. De Jager and C. De Vries, At. Data. Nucl.
Data Tables \textbf{36}, 495 (1987).

\bibitem {BBML77}G. Bertsch, J. Borysowicz, H. McManus, and W.G. Love, Nucl.
Phys. \textbf{A284}, 399 (1977).

\bibitem {Ra79}L. Ray, Phys. Rev. C 20, 1857 (1979).

\bibitem {Sag04}H. Sagawa, private communication.

\bibitem {EB02}H. Esbensen and G.F. Bertsch, Nucl. Phys. \textbf{A706}, 383 (2002).

\bibitem {Ro73}R.G.H. Robertson, Phys. Rev. \textbf{C7}, 543 (1973).

\bibitem {EB96}H. Esbensen and G.F. Bertsch, Nucl. Phys. \textbf{A600}, 600 (1996).

\bibitem {Ber96}C.A. Bertulani, Z. Phys. \textbf{A356}, 293 (1996).

\bibitem {Dav01}B. Davids, et al., Phys. Rev. \textbf{C63}, 065806 (2001).

\bibitem {Kik97}T. Kikuchi et al., Phys. Lett. \textbf{B391}, 261 (1997).

\bibitem {BG98}C.A. Bertulani and M. Gai, Nucl. Phys. \textbf{A636}, 227 (1998).

\bibitem {Sch03}F. Sch\"umann et al., Phys. Rev. Lett. \textbf{90}, 232501 (2003).

\bibitem {EB022}H. Esbensen and C.A. Bertulani, Phys. Rev. \textbf{C65},
024605 (2002).

\bibitem {EBS04}H. Esbensen, G.F. Bertsch and K. Snover, private communication and
to be published.
\end{thebibliography}
\end{document}